\documentclass[prl,twocolumn,superscriptaddress]{revtex4}
\pdfoutput=1
\usepackage{amssymb,amsmath,amsthm,graphicx}
\usepackage{graphics, color}
\usepackage{subfigure} 
\usepackage{latexsym}
\usepackage{bm}
\usepackage{epsfig}
\usepackage[pdftex]{hyperref}

\begin{document}
\title{Black holes with a single Killing vector field: \\ black resonators}
\author{\'Oscar J.~C.~Dias}
\email{O.J.Campos-Dias@soton.ac.uk}
\affiliation{STAG research centre and Mathematical Sciences, University of Southampton, UK}
\author{Jorge E.~Santos}
\email{J.E.Santos@damtp.cam.ac.uk}
\affiliation{DAMTP, Centre for Mathematical Sciences, University of Cambridge, \\ Wilberforce Road, Cambridge CB3 0WA, UK}
\author{Benson Way}
\email{B.Way@damtp.cam.ac.uk}
\affiliation{DAMTP, Centre for Mathematical Sciences, University of Cambridge, \\ Wilberforce Road, Cambridge CB3 0WA, UK}
\begin{abstract}\noindent{
We numerically construct asymptotically anti-de Sitter (AdS) black holes in four dimensions that contain only a single Killing vector field.  These solutions, which we coin \emph{black resonators}, link the superradiant instability of Kerr-AdS to the nonlinear weakly turbulent instability of AdS by connecting the onset of the superradiance instability to smooth, horizonless geometries called geons.  Furthermore, they demonstrate non-uniqueness of Kerr-AdS by sharing asymptotic charges.  Where black resonators coexist with Kerr-AdS, we find that the black resonators have higher entropy. Nevertheless, we show that black resonators are unstable and comment on the implications for the endpoint of the superradiant instability. 
}
\end{abstract}
\maketitle

\emph{\bf  Introduction.}
As the simplest of gravitating objects, black holes (BHs) play a fundamental role in our understanding of general relativity.  Indeed, four-dimensional, asymptotically flat BHs are stable and uniquely specified by their asymptotic charges \cite{Robinson:2004zz}.  However, there are circumstances where stability and uniqueness can be violated, such as those in higher dimensions \cite{Gregory:1993vy,Lehner:2010pn,Emparan:2001wn,Dias:2009iu,Shibata:2009ad,Shibata:2010wz,Dias:2010maa,Dias:2010gk,Dias:2014cia,Santos:2015iua}. We will argue that this can also be accomplished in four dimensions with asymptotically anti-de Sitter (AdS) BHs.  

Unlike Minkowski or de Sitter space, AdS contains a timelike boundary at conformal infinity where reflecting (energy and angular momentum conserving) boundary conditions are typically imposed to render the initial value problem well posed 
\cite{Friedrich:1995vb}. The presence of this boundary has drastic consequences for the stability of solutions in AdS.  For example, rotating BHs contain an ergoregion from which energy can be extracted by the Penrose process \cite{Penrose:1971uk}. For waves, this phenomenon is called superradiance \cite{Zeldovich:1971,Starobinsky:1973,Teukolsky:1974yv}.  In AdS, these waves return after scattering from the boundary and extract more energy.  The process continues until the waves contain enough energy to backreact on the geometry, causing the so-called superradiant instability \cite{Cardoso:2004hs,Dias:2013sdc,Cardoso:2013pza}.

The reflecting boundary also has implications for the stability of AdS itself.  A nonlinear instability may occur if an excitation with arbitrarily small, but finite energy around AdS continues to reflect off the boundary and eventually forms a BH.  There is numerical evidence in support of this instability with a spherically symmetric scalar field \cite{Bizon:2011gg,Buchel:2012uh,Okawa:2015xma}.  There is additionally a proposed perturbative explanation for this instability \cite{Bizon:2011gg} which applies to pure gravity and beyond spherical symmetry \cite{Dias:2011ss}.  At linear order in perturbation theory, AdS contains an infinite tower of evenly-spaced normal modes.  At higher orders, resonances between modes cause higher modes to be excited that grow linearly in time. In the generic case, this leads to a breakdown of perturbation theory, and is interpreted as the beginnings of a nonlinear instability.  This instability is called \emph{weakly turbulent} due to this energy shift from longer to shorter length scales.

Though there is a breakdown of perturbation theory for generic initial data, perturbation theory survives to arbitrarily high orders when only a single mode is excited.  This leads to a family of horizonless time-periodic solutions called \emph{oscillons} (\emph{boson stars}) for a real (complex) scalar field \cite{Bizon:2011gg,Buchel:2012uh} and \emph{geons} for pure gravity \cite{Dias:2011ss,Horowitz:2014hja}.  These geons can be thought of as nonlinear normal modes of AdS and are solutions that contain only a single Killing field.  Any gravitational radiation emitted by the geon is balanced by absorption of waves reflected from the AdS boundary.  Since perturbation theory breaks down for two-mode initial data, geons can be thought of as a basis for the nonlinear instability \cite{Dias:2011ss,Horowitz:2014hja}.  

We will construct a new family of BHs that joins the onset of the superradiant instability of Kerr-AdS \cite{Dias:2013sdc,Cardoso:2013pza}
\footnote{also known as the Carter solution.} 
to the geons.  Since these BHs are time-periodic and single out a particular frequency, we call them \emph{black resonators}.  One limit of black resonators corresponds to the onset of the superradiant instability.  Black resonators are thus the BHs predicted by \cite{Reall:2002bh} and alluded to in \cite{Kunduri:2006qa,Dias:2011ss,Cardoso:2013pza}.  The opposite, zero-size limit of black resonators corresponds to the geons where this frequency is given by a nonlinear normal mode of AdS \cite{Dias:2011ss,Horowitz:2014hja}.  Like the geons, black resonators are time-periodic and have only one Killing field.  The Killing field is also the horizon generator, so black resonators have bifurcate Killing horizons and globally well-defined horizon temperatures.
  
\emph{\bf  Numerical Approach.}
We will search for solutions to the Einstein equation with a negative cosmological constant $\Lambda=-3/L^2$.  That is, we will solve
\begin{equation}
R_{ab}+\frac{3}{L^2}g_{ab}=0\,,
\label{eq:einstein}
\end{equation}
where $L$ is the AdS length scale, $g_{ab}$ the metric and $R_{ab}$ the Ricci tensor.  We wish to find asymptotically global AdS solutions with a boundary metric conformal to the Einstein static universe
\begin{equation}
\mathrm{d}s^2_{\mathrm{bdy}} = -\mathrm{d}t^2+\mathrm{d}\theta^2+\sin^2\theta \mathrm{d}\varphi^2\,.
\label{eq:einsteinstatic}
\end{equation}

We use the DeTurck method \cite{Headrick:2009pv} which proceeds by writing down any reference metric $\bar{g}$ of our choice that shares the same causal structure as the solution $g$ we wish to find. Then, rather than solve the Einstein equation \eqref{eq:einstein}, we instead solve the Einstein-DeTurck equation:
\begin{equation}
R_{ab}+\frac{3}{L^2}-\nabla_{(a}\xi_{b)}=0\,,
\label{eq:dennis}
\end{equation}
with $\xi^a = g^{cd}[\Gamma^{a}_{cd}(g)-\Gamma^{a}_{cd}(\bar{g})]$, where $\Gamma(\mathfrak{g})$ is the Christoffel symbol associated with a metric $\mathfrak{g}$.  Since we are solving different equations, we verify {\it a posteriori} that our solutions to \eqref{eq:dennis} satisfies $\xi_a=0$ to machine precision, and hence is also a solution to \eqref{eq:einstein}.  In some cases, it is possible to prove that all solutions of \eqref{eq:dennis} must have $\xi_a=0$ \cite{Figueras:2011va}, but we do not have such a proof for our case of interest.  However, we have verified that \eqref{eq:dennis} yields elliptic partial differential equations for which local uniqueness theorems exist \cite{Adam:2011dn}.  Thus, solutions with $\xi\neq0$ cannot have $\xi$ arbitrarily small for all ranges of parameters, and are therefore distinguishable from solutions with $\xi=0$.

Following the DeTurck method, we need an ansatz and a reference metric containing a single Killing field,
\begin{equation}
K=\partial_t+\Omega_H \partial_\varphi\,,
\end{equation}
where $\Omega_H$ will be the horizon angular velocity. $K$ is a Killing field, but $\partial_t$ and $\partial_\varphi$ are not. Thus, the solution will neither be time independent nor axisymmetric, but is instead time-periodic. We now perform the following change of variables: $\mathrm{d}\tau = \mathrm{d}t$ and $\mathrm{d}\phi=\mathrm{d}\varphi+\Omega_H \mathrm{d}t$. In these coordinates, $K=\partial_\tau$ and the boundary metric is:
\begin{equation}
\mathrm{d}s^2_{\mathrm{bdy}} = -\mathrm{d}\tau^2+\mathrm{d}\theta^2+\sin^2\theta (\mathrm{d}\phi-\Omega_H \mathrm{d}\tau)^2\,.
\end{equation}

For a general ansatz containing a Killing horizon generated by $K$, we choose:
\begin{widetext}
\begin{multline}
\mathrm{d}s^2 = \frac{L^2}{(1-y^2)^2}\Bigg[-y^2 q_1\Delta(y)\left(\mathrm{d}\tau+y\,q_6\mathrm{d}y\right)^2+\frac{4y_+^2\,q_2\mathrm{d}y^2}{\Delta(y)}
+\frac{4y_+^2\,q_3}{2-x^2}\left(\mathrm{d}x+yx\sqrt{2-x^2}\,q_7\mathrm{d}y+y^2x\sqrt{2-x^2}\,q_8 \mathrm{d}\tau\right)^2\\
+(1-x^2)^2y_+^2 q_4\left(\mathrm{d}\phi-y^2 q_5 \mathrm{d}\tau+\frac{x\sqrt{2-x^2}q_9\mathrm{d}x}{1-x^2}+y\,q_{10}\mathrm{d}y\right)^2\Bigg]\,,
\label{eq:bigmess}
\end{multline}
\end{widetext}
with $\Delta(y) = (1-y^2)^2+y_+^2(3-3y^2+y^4)$ and $q_i=q_i(x,y,\phi)$. As a reference metric, we take \eqref{eq:bigmess} with $q_1=q_2=q_3=q_4=1$, $q_5 = \Omega_H$ and $q_i=0$, for $i\geq 6$. 

If $\Omega_H=0$, our reference metric is the Schwarzschild-AdS BH with entropy $S=\pi y_+^2 L^2$ and temperature
\begin{equation}
T=\frac{1+3y_+^2}{4\pi y_+ L}\,.
\label{eq:temperature}
\end{equation}
To recover more familiar coordinates, redefine $r/L = y_+/(1-y^2)$ and $\cos \theta = x\sqrt{2-x^2}$.  We chose this reference metric rather than Kerr-AdS for tidiness.

Since black resonators branch from the onset of the superradiant instability of Kerr-AdS \cite{Dias:2013sdc,Cardoso:2013pza}, let us now describe this instability in more detail.  Perturbations of Kerr-AdS are labeled by a type (scalar or vector) and polar and azimutal wavenumbers $\ell$ and $m$, respectively.  While all Kerr-AdS BHs with $\Omega_HL>1$ are superradiant unstable, different perturbative modes have onsets in difference places in parameter space.  At these onsets are zero modes which mark the appearance of a new family of solutions which will be the black resonators.  

For definitiveness and simplicity, we will focus on the scalar mode $\ell = m=2$, which has an extra discrete symmetry under $x\to-x$.  The Kerr-AdS parameters corresponding to the onset of the superradiant instability of this mode were obtained in \cite{Dias:2013sdc,Cardoso:2013pza} 
\footnote{The onset of superradiant modes with oppositely signed amplitudes yield the same physical black resonators.}.

Let us now discuss boundary conditions.  At $y=0$, we demand a regular bifurcate Killing horizon generated by $\partial_\tau$ with temperature \eqref{eq:temperature}. This amounts to $q_1(x,0,\phi)=q_2(x,0,\phi)$, with the remaining functions having Neumann conditions.  The boundary ($y=1$) must approach that of the reference metric, which is a Dirichlet condition.  The discrete symmetry $x\to-x$ requires that all of the $q_i$'s have Neumann boundary conditions at $x=0$.  At $x=1$, regularity requires $q_7(1,y,\phi)=q_8(1,y,\phi)=0$ and Neumann conditions for the remaining functions. Finally, $m=2$ requires that $\phi$ be periodic in $\phi\in(0,\pi]$.

With equations and boundary conditions, we can now solve the system of partial differential equations by numerical methods.  We use a standard Newton-Raphson algorithm and discretise the Einstein-DeTurck equations using pseudospectral collocation (Chebyshev-Gauss-Lobatto nodes along the $x$ and $y$ directions, and Fourier nodes along the $\phi$ direction).  The resulting algebraic linear systems are solved by LU decomposition. 

Our reference metric is not Kerr-AdS, so we must construct Kerr-AdS numerically.  We fix $y_+$, assume independence in $\phi$, and slowly increase $\Omega_H$ from zero until we are at the onset of the superradiant instability.  Note that this is consistent with our boundary conditions above.  

From here, we have tried to obtain black resonators by varying $\Omega_H$ and $y_+$ using a perturbed Kerr-AdS as a seed.  This proved unsuccessful since Kerr-AdS is too strong an attractor for Newton-Raphson.  Instead, we used $y_+$ and a \emph{wiggliness} parameter $\varepsilon$ defined as
\begin{equation}
\varepsilon \equiv \int_{0}^{\pi}q_9(1,0,\phi)\sin (m\,\phi)\mathrm{d}\phi \,.
\label{eq:wiggliness}
\end{equation}
We add the defining equation \eqref{eq:wiggliness} to our system of equations and $\Omega_H$ as an extra unknown.  Any solution with $\varepsilon\neq0$ has $\phi$ dependence and hence cannot be Kerr-AdS. 

\begin{figure}
\centering
\includegraphics[height = 0.32\textheight]{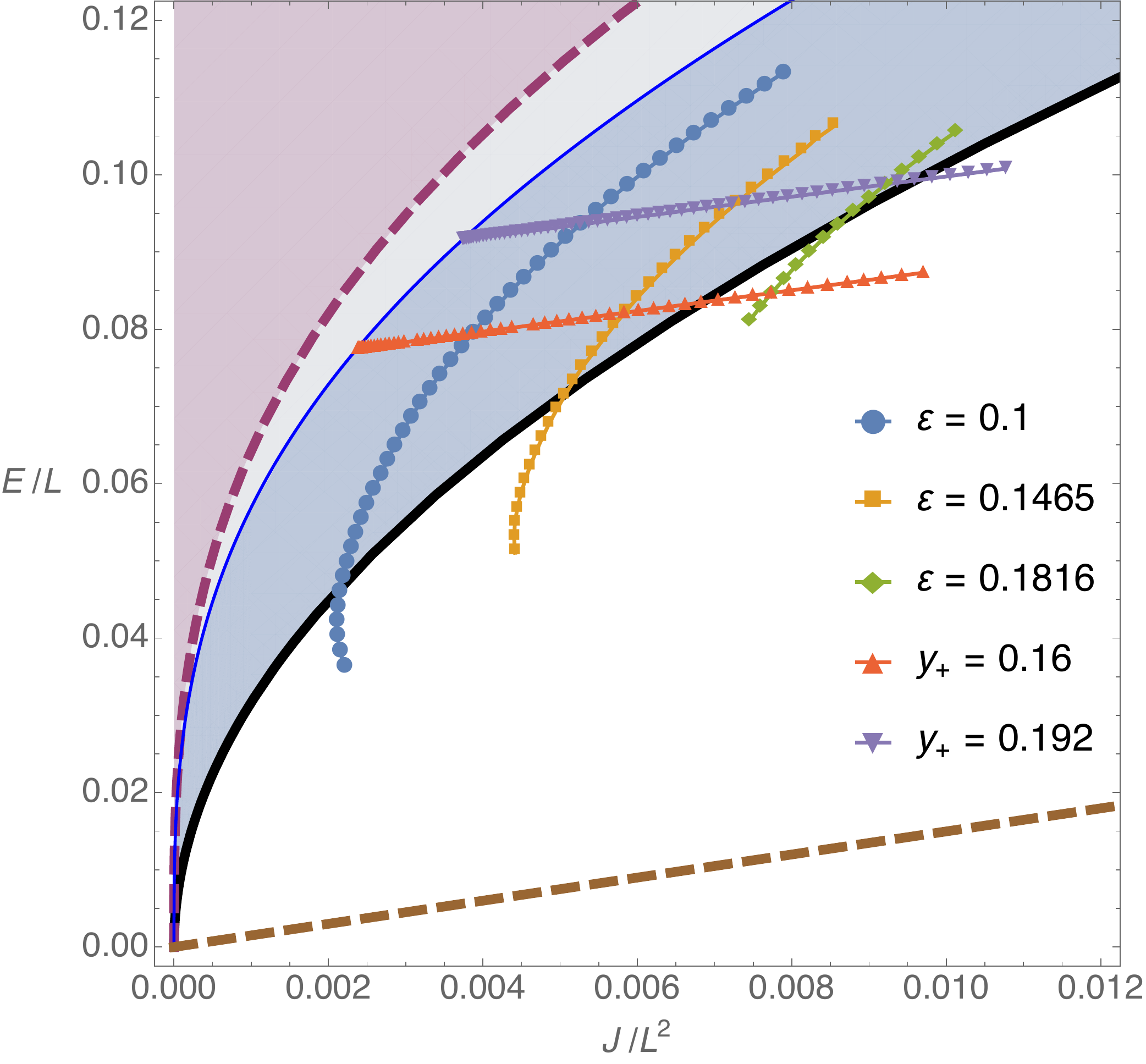}
\caption{\label{fig:1} $E$ vs. $J$ phase diagram of AdS black holes with Kerr-AdS BHs (all regions above thick solid black line), geons in \cite{Horowitz:2014hja} (bottom yellow dashed line), and  black resonators (data points).  Also plotted are  Kerr-AdS BHs with $\Omega_H L \leq 1$ (dashed purple line and purple region above it), extremal Kerr-AdS BHs (thick solid black line), and the onset of scalar $m=\ell=2$ mode in \cite{Cardoso:2013pza} (thin blue line).}
\end{figure}

\emph{\bf Results.}
Our main result is shown in Fig.~\ref{fig:1}, where the energy ($E$) versus angular momentum ($J$) phase diagram of rotating BHs in AdS is presented in units of the AdS radius $L$.  Kerr-AdS BHs lie above the thick solid black line, which refers to extremal ($T=0$) Kerr-AdS BHs.  Kerr-AdS BHs with $\Omega_H L<1$ are likely linearly stable \cite{Hawking:1999dp,Cardoso:2013pza} and lie in upper-left region above the dashed purple line (the purple line itself refers to $\Omega_H L=1$).  The onset of the superradiant instability for the scalar $\ell=m=2$ modes \cite{Cardoso:2013pza} lie on the thin blue line.  The geons from \cite{Horowitz:2014hja} lie on the bottom dashed line.  The data points refer to the black resonators we have constructed, with parameters indicated by the plot markers.  

From Fig.~\ref{fig:1}, we see that our solutions at small $\varepsilon$ recover the onset curve of \cite{Cardoso:2013pza}. This is a strong consistency check since the onset curve was generated using the Teukolsky equation in AdS, which only indirectly uses the metric \cite{Dias:2013sdc}.  All of our black resonators also have $\Omega_H L>1$, which indicates that no Killing vector can be found near the conformal boundary that is everywhere timelike.  We also see that some black resonators extend below the extremal limit of Kerr-AdS BHs and are therefore the only known regular solutions with these asymptotic charges. 

\begin{figure}
\centering
\includegraphics[height = 0.2\textheight]{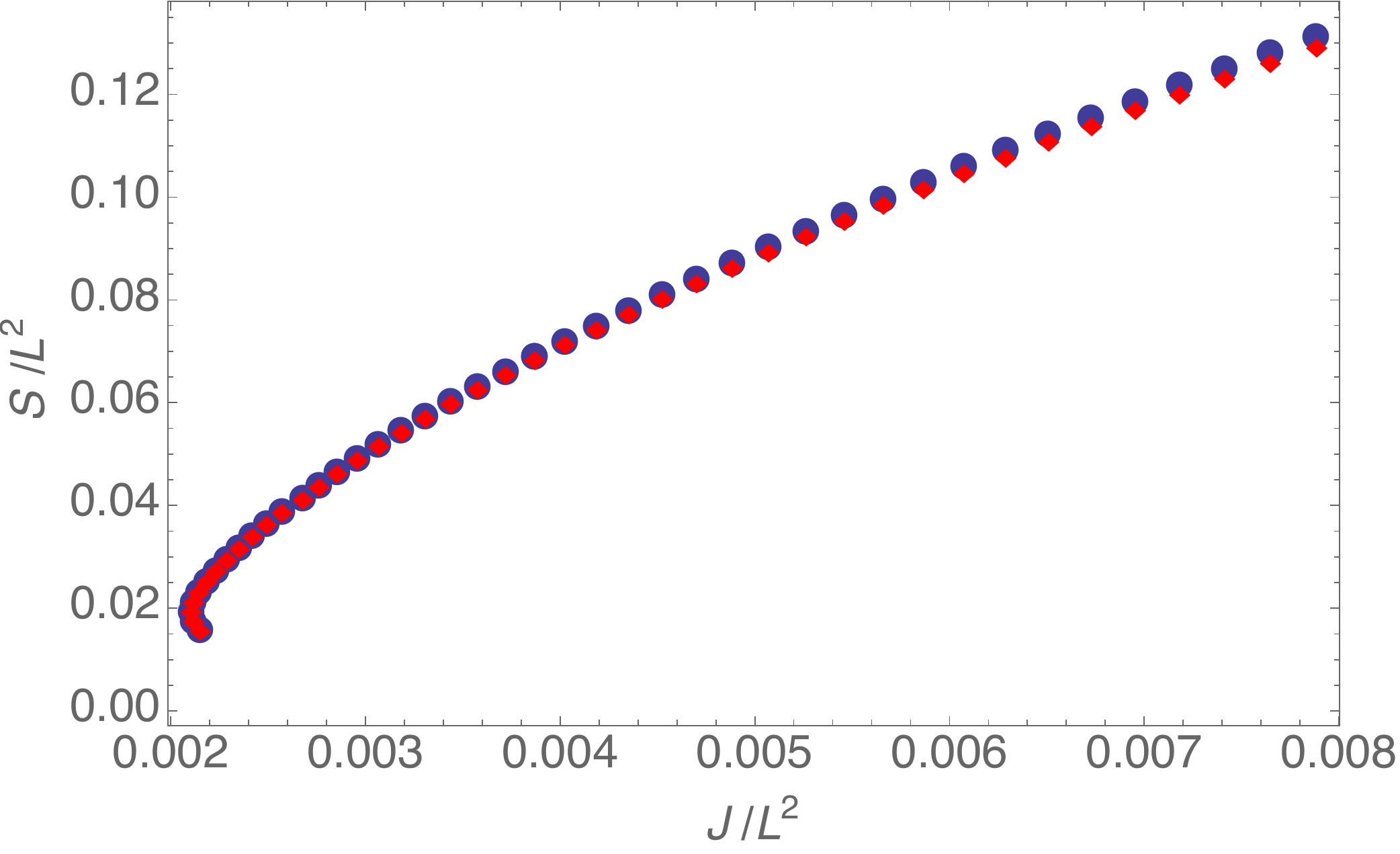}
\caption{\label{fig:thermomodel} Comparison of numerical $\varepsilon=0.1$ black resonators (blue disks) and the perturbative result \eqref{Spert} (red diamonds).}
\end{figure}

Though we focused on the scalar $\ell=m=2$ mode, a perturbative expansion can predict the entropy of small $E$ and $J$ black resonators for more general modes \cite{Cardoso:2013pza}.  For scalar modes with $\ell=m$, the entropy is
\begin{equation}\label{Spert}
S = 4\pi E^2\left[1-\left(1+\frac{1}{m}\right)\frac{J}{E\,L}\right]^2.
\end{equation}
For small black resonators, Fig.~\ref{fig:thermomodel} shows good agreement between our numerical results and  the perturbative prediction of  \eqref{Spert} with $m=2$.  Since \eqref{Spert} assumes that zero-size black resonators merge with geons, this agreement can be taken as evidence that black resonators connect to geons.  Fig.~\ref{fig:1} shows black resonators approaching geons, but only down to size $y_+\sim 0.07$.

We note that if black resonators with arbitrary $m$ are connected to geons, the minimum $E$ for fixed $J$ that black resonators can have occurs when $m\to+\infty$ for an arbitrarily small black resonator (\emph{i.e.} a geon with $m=+\infty$).  This configuration saturates the BPS bound $E=J/L$.

\begin{figure}
\centering
\includegraphics[height = 0.2\textheight]{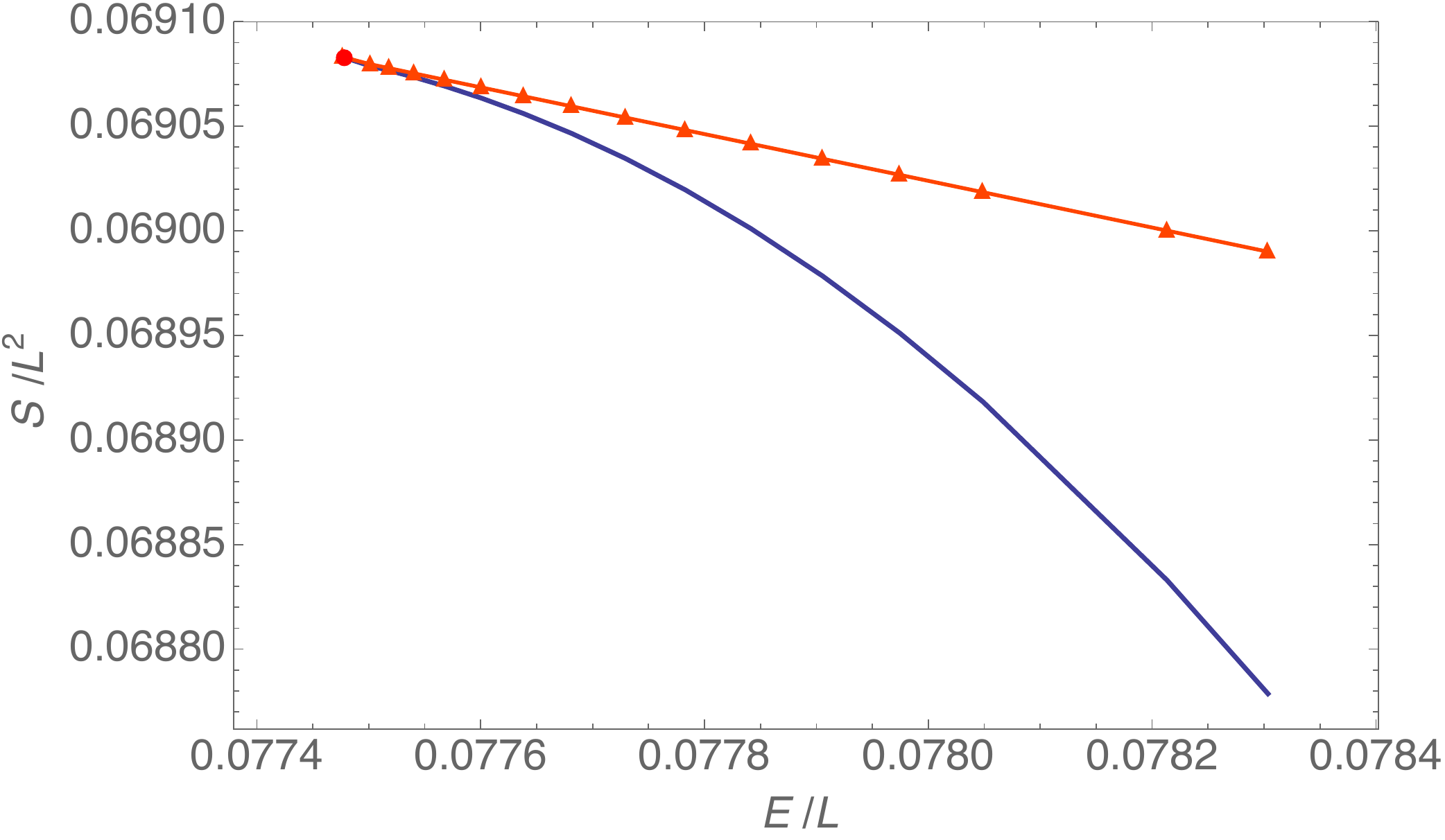}
\caption{\label{fig:2} Entropy of rotating BHs in AdS versus their energy with Kerr-AdS BHs (solid blue line) and black resonators (red triangles).  The black resonators all have $y_+=0.16$.}
\end{figure}

Another quantity of interest is the entropy of black resonators compared with that of Kerr-AdS.  At the same asymptotic charges $E$ and $J$, we find that black resonators have higher entropy than Kerr-AdS ( see for instance Fig.~\ref{fig:2}).  These solutions merge at the onset of superradiance through a second order phase transition. 

\begin{figure}
\centering
\includegraphics[height = 0.24\textheight]{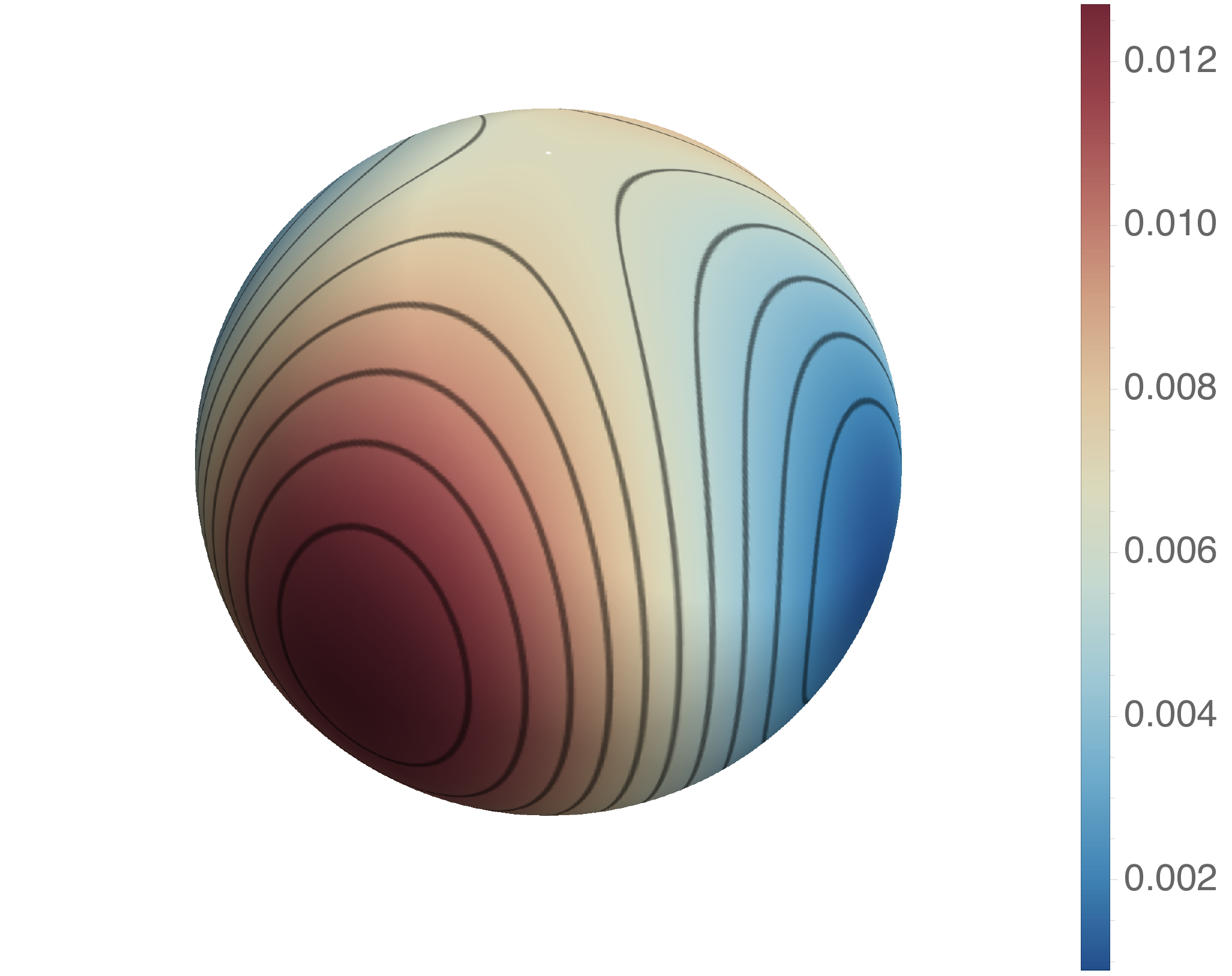}
\caption{\label{fig:3} Energy density in units of $L$ at a moment in time of a black resonator with $y_+=0.16$ and $\varepsilon=0.1$.}
\end{figure}

We show the energy density on the $S^2$ of the boundary metric \eqref{eq:einsteinstatic} in Fig.~\ref{fig:3} \cite{deHaro:2000xn}. This figure represents an instant in time and should be imagined as rotating with angular velocity $\Omega_H$ (\emph{i.e.} in a time-periodic way).

\emph{\bf Outlook.}
We have constructed new BHs in AdS with a single Killing field which we call black resonators \footnote{Similar black resonators may also exist inside Dirichlet walls with $E$ and $J$ conserving boundary conditions.}. Black resonators are not ruled out by Hawking's rigidity theorem \cite{Hawking:1971vc,Hollands:2006rj,Moncrief:2008mr,Hollands:2012xy} since the single Killing field also generates the horizon. These BHs branch from the onset of the superradiant instability in Kerr-AdS, and extend in their zero-size limit to smooth horizonless geometries called geons \cite{Dias:2011ss,Horowitz:2014hja}. The existence of black resonators proves that Kerr-AdS is non-unique, even in four dimensions. We focused on the scalar progenitor mode $m=\ell=2$, but expect similar behaviour for other $m$ and $\ell$. This would mean a countably infinite violation of uniqueness for rotating BHs in AdS with $\Omega_H L>1$ and $E>J/L$.  

In retrospect, new BHs could have been anticipated from the AdS/CFT correspondence.  Since CFTs are expected to saturate the bound $E\geq J/L$, but Kerr-AdS BHs do not, another BH might fill the gap. Though, this argument does not suggest that these BHs have a single Killing field or are connected to the superradiant onset.

The precise boundary CFT interpretation of these instabilities and the black resonators remain mysterious.  We note that superradiance is not particular to four dimensions, and occurs also in $AdS_5$.  Furthermore, including the full $AdS_5\times S^5$ bulk geometry, so that the boundary field theory is specifically $\mathcal N=4$ super Yang-Mills, does not cure this instability.  

Though these black resonators have more entropy than Kerr-AdS, we argue that they are unstable, so they cannot be the endpoint of the superradiant instability. While black resonators with progenitor modes $\ell=m=2$ should be stable to perturbations with $m=\ell=2$, they are unstable to higher $m$ modes. The reason for this is that small black resonators are well-approximated (as confirmed in \eqref{Spert} and Fig. \ref{fig:thermomodel}) by a small Kerr-AdS BH at the centre of a geon \cite{Dias:2011ss,Cardoso:2013pza}, and small Kerr-AdS BHs are still unstable to higher $m$ modes.  More precisely, the results of \cite{wald1} mathematically prove that our solutions are unstable, since no Killing vector field that is everywhere timelike can be found at the conformal boundary.

Our results support the conjecture of \cite{Dias:2011at} that there is no stationary endpoint to the superradiant instability in AdS. Instead, modes with increasing $m$ continue to be excited and develop.  It may be possible for additional small BHs to form as energy is deposited into higher $m$ modes. While such configurations should exist, they are themselves superradiantly unstable \cite{Dias:2012tq,santososcar1}. Though classical evolution may continue indefinitely, eventually the increasingly high $m$ modes will reach sub-Planckian length scales.  This may be viewed as a violation of the spirit of cosmic censorship in that initial data well-described classically leads to a situation requiring quantum mechanics.  

\begin{acknowledgments}
\begin{center}
\emph{\bf  Acknowledgments}
\end{center}
It is a pleasure to thank Dave Berenstein and Don Marolf for discussions. The authors thankfully acknowledge the computer resources, technical expertise, and assistance provided by CENTRA/IST. Some computations were performed at the cluster `Baltasar-Sete-S\'ois', supported by the DyBHo-256667 ERC Starting Grant; and on the COSMOS Shared Memory system at DAMTP, University of Cambridge operated on behalf of the STFC DiRAC HPC Facility and funded by BIS National E-infrastructure capital grant ST/J005673/1 and STFC grants ST/H008586/1, ST/K00333X/1. O.J.C.D. is supported by the STFC Ernest Rutherford grants ST/K005391/1 and ST/M004147/1. This research received funding from the European Research Council under the European Community's 7th Framework Programme (FP7/2007-2013)/ERC grant agreement no. [247252]. B.W. is supported by European Research Council grant no. ERC-2011-StG 279363-HiDGR.
\end{acknowledgments}


\bibliography{refs}{}
\bibliographystyle{JHEP}

\end{document}